\documentclass[aps,prl,twocolumn,showpacs,groupedaddress,floatfix]{revtex4}  
\usepackage{graphicx}  
\usepackage{dcolumn}   
\usepackage{bm}        
\usepackage{amssymb}   

\newcommand{\met}{\mbox{$\not\!\!E_T$}}
\newcommand{\xmet}{\mbox{$\not\!\!E'_T$}}
\newcommand{\whprocess}{
\mbox{$WH\rightarrow WWW^*\rightarrow$}
\mbox{$l^\pm\nu\ l'^\pm\nu'+X$}}
\newcommand{\wzprocess}{\mbox{$WZ\rightarrow l\nu\ l'l'$}}
\newcommand{\zzprocess}{\mbox{$ZZ\rightarrow ll\ l'l'$}}
\newcommand{\wwprocess}{\mbox{$WW\rightarrow l\nu\ l'\nu'$}}

\begin{document}


\hspace{5.2in} \mbox{Fermilab-Pub-06-246-E}

\title{Search for associated Higgs boson production
\whprocess\\
in $p\bar{p}$ collisions at $\sqrt{s}=$1.96~TeV
}
%
\author{                                                                      
V.M.~Abazov,$^{36}$                                                           
B.~Abbott,$^{76}$                                                             
M.~Abolins,$^{66}$                                                            
B.S.~Acharya,$^{29}$                                                          
M.~Adams,$^{52}$                                                              
T.~Adams,$^{50}$                                                              
M.~Agelou,$^{18}$                                                             
J.-L.~Agram,$^{19}$                                                           
S.H.~Ahn,$^{31}$                                                              
M.~Ahsan,$^{60}$
% ...
\\                                                                            
\vskip 0.30cm                                                                 
\centerline{(D\O\ Collaboration)}                                             
\vskip 0.30cm                                                                 
}                                                                             
\affiliation{                                                                 
\centerline{$^{1}$Universidad de Buenos Aires, Buenos Aires, Argentina}       
\centerline{$^{2}$LAFEX, Centro Brasileiro de Pesquisas F{\'\i}sicas,         
                  Rio de Janeiro, Brazil}                                     
\centerline{$^{3}$Universidade do Estado do Rio de Janeiro,                   
                  Rio de Janeiro, Brazil}                                     
\centerline{$^{4}$Instituto de F\'{\i}sica Te\'orica, Universidade            
                  Estadual Paulista, S\~ao Paulo, Brazil}                     
\centerline{$^{5}$University of Alberta, Edmonton, Alberta, Canada,           
                  Simon Fraser University, Burnaby, British Columbia, Canada,}
\centerline{York University, Toronto, Ontario, Canada, and                    
                  McGill University, Montreal, Quebec, Canada}                
\centerline{$^{6}$Institute of High Energy Physics, Beijing,                  
                  People's Republic of China}                                 
\centerline{$^{7}$University of Science and Technology of China, Hefei,       
                  People's Republic of China}                                 
\centerline{$^{8}$Universidad de los Andes, Bogot\'{a}, Colombia}             
\centerline{$^{9}$Center for Particle Physics, Charles University,            
                  Prague, Czech Republic}                                     
\centerline{$^{10}$Czech Technical University, Prague, Czech Republic}        
\centerline{$^{11}$Center for Particle Physics, Institute of Physics,         
                   Academy of Sciences of the Czech Republic,                 
                   Prague, Czech Republic}                                    
\centerline{$^{12}$Universidad San Francisco de Quito, Quito, Ecuador}        
\centerline{$^{13}$Laboratoire de Physique Corpusculaire, IN2P3-CNRS,         
                   Universit\'e Blaise Pascal, Clermont-Ferrand, France}      
\centerline{$^{14}$Laboratoire de Physique Subatomique et de Cosmologie,      
                   IN2P3-CNRS, Universite de Grenoble 1, Grenoble, France}    
\centerline{$^{15}$CPPM, IN2P3-CNRS, Universit\'e de la M\'editerran\'ee,     
                   Marseille, France}                                         
\centerline{$^{16}$IN2P3-CNRS, Laboratoire de l'Acc\'el\'erateur              
                   Lin\'eaire, Orsay, France}                                 
\centerline{$^{17}$LPNHE, IN2P3-CNRS, Universit\'es Paris VI and VII,         
                   Paris, France}                                             
\centerline{$^{18}$DAPNIA/Service de Physique des Particules, CEA, Saclay,    
                   France}                                                    
\centerline{$^{19}$IPHC, IN2P3-CNRS, Universit\'e Louis Pasteur, Strasbourg,  
                    France, and Universit\'e de Haute Alsace,                 
                    Mulhouse, France}                                         
\centerline{$^{20}$Institut de Physique Nucl\'eaire de Lyon, IN2P3-CNRS,      
                   Universit\'e Claude Bernard, Villeurbanne, France}         
\centerline{$^{21}$III. Physikalisches Institut A, RWTH Aachen,               
                   Aachen, Germany}                                           
\centerline{$^{22}$Physikalisches Institut, Universit{\"a}t Bonn,             
                   Bonn, Germany}                                             
\centerline{$^{23}$Physikalisches Institut, Universit{\"a}t Freiburg,         
                   Freiburg, Germany}                                         
\centerline{$^{24}$Institut f{\"u}r Physik, Universit{\"a}t Mainz,            
                   Mainz, Germany}                                            
\centerline{$^{25}$Ludwig-Maximilians-Universit{\"a}t M{\"u}nchen,            
                   M{\"u}nchen, Germany}                                      
\centerline{$^{26}$Fachbereich Physik, University of Wuppertal,               
                   Wuppertal, Germany}                                        
\centerline{$^{27}$Panjab University, Chandigarh, India}                      
\centerline{$^{28}$Delhi University, Delhi, India}                            
\centerline{$^{29}$Tata Institute of Fundamental Research, Mumbai, India}     
\centerline{$^{30}$University College Dublin, Dublin, Ireland}                
\centerline{$^{31}$Korea Detector Laboratory, Korea University,               
                   Seoul, Korea}                                              
\centerline{$^{32}$SungKyunKwan University, Suwon, Korea}                     
\centerline{$^{33}$CINVESTAV, Mexico City, Mexico}                            
\centerline{$^{34}$FOM-Institute NIKHEF and University of                     
                   Amsterdam/NIKHEF, Amsterdam, The Netherlands}              
\centerline{$^{35}$Radboud University Nijmegen/NIKHEF, Nijmegen, The          
                  Netherlands}                                                
\centerline{$^{36}$Joint Institute for Nuclear Research, Dubna, Russia}       
\centerline{$^{37}$Institute for Theoretical and Experimental Physics,        
                   Moscow, Russia}                                            
\centerline{$^{38}$Moscow State University, Moscow, Russia}                   
\centerline{$^{39}$Institute for High Energy Physics, Protvino, Russia}       
\centerline{$^{40}$Petersburg Nuclear Physics Institute,                      
                   St. Petersburg, Russia}                                    
\centerline{$^{41}$Lund University, Lund, Sweden, Royal Institute of          
                   Technology and Stockholm University, Stockholm,            
                   Sweden, and}                                               
\centerline{Uppsala University, Uppsala, Sweden}                              
\centerline{$^{42}$Physik Institut der Universit{\"a}t Z{\"u}rich,            
                   Z{\"u}rich, Switzerland}                                   
\centerline{$^{43}$Lancaster University, Lancaster, United Kingdom}           
\centerline{$^{44}$Imperial College, London, United Kingdom}                  
\centerline{$^{45}$University of Manchester, Manchester, United Kingdom}      
\centerline{$^{46}$University of Arizona, Tucson, Arizona 85721, USA}         
\centerline{$^{47}$Lawrence Berkeley National Laboratory and University of    
                   California, Berkeley, California 94720, USA}               
\centerline{$^{48}$California State University, Fresno, California 93740, USA}
\centerline{$^{49}$University of California, Riverside, California 92521, USA}
\centerline{$^{50}$Florida State University, Tallahassee, Florida 32306, USA} 
\centerline{$^{51}$Fermi National Accelerator Laboratory,                     
            Batavia, Illinois 60510, USA}                                     
\centerline{$^{52}$University of Illinois at Chicago,                         
            Chicago, Illinois 60607, USA}                                     
\centerline{$^{53}$Northern Illinois University, DeKalb, Illinois 60115, USA} 
\centerline{$^{54}$Northwestern University, Evanston, Illinois 60208, USA}    
\centerline{$^{55}$Indiana University, Bloomington, Indiana 47405, USA}       
\centerline{$^{56}$University of Notre Dame, Notre Dame, Indiana 46556, USA}  
\centerline{$^{57}$Purdue University Calumet, Hammond, Indiana 46323, USA}    
\centerline{$^{58}$Iowa State University, Ames, Iowa 50011, USA}              
\centerline{$^{59}$University of Kansas, Lawrence, Kansas 66045, USA}         
\centerline{$^{60}$Kansas State University, Manhattan, Kansas 66506, USA}     
\centerline{$^{61}$Louisiana Tech University, Ruston, Louisiana 71272, USA}   
\centerline{$^{62}$University of Maryland, College Park, Maryland 20742, USA} 
\centerline{$^{63}$Boston University, Boston, Massachusetts 02215, USA}       
\centerline{$^{64}$Northeastern University, Boston, Massachusetts 02115, USA} 
\centerline{$^{65}$University of Michigan, Ann Arbor, Michigan 48109, USA}    
\centerline{$^{66}$Michigan State University,                                 
            East Lansing, Michigan 48824, USA}                                
\centerline{$^{67}$University of Mississippi,                                 
            University, Mississippi 38677, USA}                               
\centerline{$^{68}$University of Nebraska, Lincoln, Nebraska 68588, USA}      
\centerline{$^{69}$Princeton University, Princeton, New Jersey 08544, USA}    
\centerline{$^{70}$State University of New York, Buffalo, New York 14260, USA}
\centerline{$^{71}$Columbia University, New York, New York 10027, USA}        
\centerline{$^{72}$University of Rochester, Rochester, New York 14627, USA}   
\centerline{$^{73}$State University of New York,                              
            Stony Brook, New York 11794, USA}                                 
\centerline{$^{74}$Brookhaven National Laboratory, Upton, New York 11973, USA}
\centerline{$^{75}$Langston University, Langston, Oklahoma 73050, USA}        
\centerline{$^{76}$University of Oklahoma, Norman, Oklahoma 73019, USA}       
\centerline{$^{77}$Oklahoma State University, Stillwater, Oklahoma 74078, USA}
\centerline{$^{78}$Brown University, Providence, Rhode Island 02912, USA}     
\centerline{$^{79}$University of Texas, Arlington, Texas 76019, USA}          
\centerline{$^{80}$Southern Methodist University, Dallas, Texas 75275, USA}   
\centerline{$^{81}$Rice University, Houston, Texas 77005, USA}                
\centerline{$^{82}$University of Virginia, Charlottesville,                   
            Virginia 22901, USA}                                              
\centerline{$^{83}$University of Washington, Seattle, Washington 98195, USA}  
}                                                                             
\date{October 13, 2006}

\begin{abstract}
We present a search for associated Higgs boson production in the
process $p\bar{p}\,\rightarrow WH\rightarrow WWW^*\rightarrow
l^\pm\nu\ l'^\pm\nu'+X$ in final states containing two like-sign
isolated electrons or muons ($e^\pm e^\pm$, $e^\pm \mu^\pm$, or
$\mu^\pm \mu^\pm$). The search is based on D0 Run~II data samples
corresponding to integrated luminosities of 360--380~pb$^{-1}$. No
excess is observed over the predicted standard model background. We
set 95\% C.L. upper limits on
\mbox{$\sigma(p\bar{p}\rightarrow WH)\times Br(H\rightarrow WW^*)$}
between 3.2 and 2.8~pb for Higgs boson masses from 115 to 175~GeV.
\end{abstract}

\maketitle 

The Higgs boson $H$ is a hypothesized particle introduced in the
standard model (SM) that provides the mechanism by which particles
acquire mass. While Higgs boson searches in the low mass region focus
on the $H\rightarrow b\bar{b}$ decay mode, the $H\rightarrow WW^*$
decay mode dominates for SM Higgs boson masses above 135~GeV~\cite{hdecays}.
Furthermore, in some models with anomalous couplings (``fermiophobic
Higgs boson''), the branching fraction $Br(H\rightarrow W^{(*)}W^*)$ may be
close to 100\% for Higgs masses as low as
$\approx$100~GeV~\cite{fermiophobic-higgs}.

In this Letter, we present a search for associated Higgs boson
production ($p\bar{p}\rightarrow WH$) where the Higgs boson decays
into a $WW^*$ pair, and each of the two $W$ bosons with the same
charge decay to a charged lepton (electron or muon) plus a
neutrino. The final state is characterized by two like-sign, high
transverse momentum ($p_T$), isolated charged leptons and missing
transverse energy ($\met$) due to escaping neutrinos. This decay mode
is easier to detect than $H\rightarrow b\bar{b}$ since the latter
suffers from a large irreducible $Wb\bar{b}$ background. The presence
of two like-sign leptons from $W$ decays makes this channel
advantageous over direct Higgs production,
\mbox{$p\bar{p}\rightarrow H\rightarrow WW^*$},
where the two leptons from $W$ decays have opposite signs, resulting
in large SM backgrounds ($Z/\gamma^*$, $WW$, and $t\bar{t}$
production). The main physics background in our case is \wzprocess,
and at a much lower rate, \zzprocess. The irreducible physics
background, non-resonant triple vector boson production ($VVV$,
$V=W,Z$), has a cross section that is much lower than the signal one,
as does $t\bar{t}+V$.

We use data collected by the D0 detector at the Fermilab Tevatron
Collider between April 2002 and August 2004. That data sample
corresponds to 380~pb$^{-1}$ of integrated luminosity in the $ee$
channel, 370~pb$^{-1}$ in the $e\mu$ channel, and 360~pb$^{-1}$ in the
$\mu\mu$ channel, with the variations related primarily to different
trigger requirements.

The D0 detector is described in detail elsewhere~\cite{d0det}. Its
principal elements are a central-tracking system embedded in a 2~T
superconducting solenoidal magnet, a liquid-argon/uranium calorimeter,
and an outer muon system. The central-tracking system consists of a
silicon microstrip tracker (SMT) and a central fiber tracker (CFT)
that provide tracking and vertexing for pseudorapidities $|\eta|<3$
and $|\eta|<2.5$, respectively~\cite{etadef}. The calorimeter has a
central section (CC) covering $|\eta|<$1.1, and two end calorimeters
that extend coverage to $|\eta|\approx 4.2$. The outer muon system, at
$|\eta|<2$, consists of a layer of tracking detectors and
scintillation trigger counters in front of 1.8~T iron toroids,
followed by two similar layers after the toroids.

The signal candidate events are selected by dilepton triggers.
Offline, the electrons are reconstructed as clusters in the
electromagnetic part of the CC with central tracks pointing to
them. The electron energy is measured in the calorimeter, and the
tracks provide measurement of the direction and charge. The selected
electromagnetic cluster candidates must be isolated in the calorimeter,
have a
longitudinal and transverse shower shape consistent with that of an
electron, and pass a likelihood requirement that includes a spatial
and momentum match between the cluster and the track, the electron
track isolation, and other quantities. The muons are reconstructed in
the outer muon system and matched to central tracks, their momenta
being measured in the central-tracking system. They are required to be
isolated, which means the minimum distance to the nearest jet in the
event $\Delta R(\mu,{\rm j})$~\cite{drdef} is greater than 0.5, and
the scalar sum of the transverse momenta of tracks in the $\Delta
R<0.5$ cone around the muon track (excluding this track) is less than
4~GeV. Both electrons and muons are required to have transverse
momenta greater than 15~GeV.

The efficiency for \whprocess\ signal events to pass the selection was
calculated using the {\sc pythia}~6.2~\cite{pythia} event generator
followed by a detailed simulation of the D0 detector based on the
{\sc geant}~\cite{geant} package. We use the simulation to obtain the
total acceptance and apply trigger and reconstruction efficiencies
derived from the data. The same approach was used to simulate
backgrounds from
\wzprocess\ and \zzprocess. These backgrounds are normalized to their
next-to-leading-order cross sections calculated by the
{\sc mcfm}~\cite{mcfm} program using the CTEQ6.1M parton distribution
functions~\cite{cteq6}.

In addition to the physics backgrounds mentioned above, there are two
types of instrumental background. One type, referred to as ``charge
flips,'' originates from the misreconstruction of the charge of one of
the leptons. For the same lepton flavor channels ($ee$ and $\mu\mu$)
this background is dominated by $Z/\gamma^*\rightarrow ll$. The second
type of background is like-sign lepton pairs from multijet or $W$+jets
production. In the case of muons, these can be real muons from
semileptonic heavy flavor decays that pass the isolation cuts,
punch-through hadrons misidentified as muons, or muons from $\pi/K$
decays in flight. In the case of electrons, the background originates
from electrons in semileptonic heavy flavor decays and from $\gamma$
conversions or from hadrons misidentified as electrons. This second
type of background will be referred to as ``QCD.''  There are other
processes which are included in these two background categories. In
particular, charge flips include events due to \wwprocess\ production
where one lepton charge is mismeasured. The decay $t\bar{t}\rightarrow
ll'+X$ may contribute to either charge flips (if one of the lepton
charges is mismeasured) or QCD (if a lepton from a semileptonic
$b$~decay passes the lepton identification cuts). The decay
$t\bar{t}\rightarrow l$+jets with a lepton from $b$~decay may
contribute to QCD background.

In order to reduce instrumental backgrounds, tighter track selection
is needed. The lepton tracks are required to have at least 2 (out of
an average of 8) SMT measurements and at least 5 (out of 16 possible)
CFT measurements. Also, they must originate from the primary vertex,
which is achieved by requiring the distance between the track origin
and primary vertex along the beam to be less than 1~cm, the distance
of closest approach (DCA) to the primary vertex in the transverse
plane to be less than 0.1~cm, and the DCA significance (DCA divided by
its uncertainty) to be less than 3. These cuts suppress both
charge flip (due to improved track quality) and QCD (which is enriched
with secondary leptons from $b$~decays) backgrounds.

After all these selections, we are left with a sample of 15 $ee$, 7
$e\mu$, and 12 $\mu\mu$ events, still dominated by instrumental
background. In order to further improve the signal-to-background
ratio, we perform a final selection based on a topological likelihood
discriminant
\mbox{TLD$=\prod_i s_i/(\prod_i s_i + \prod_i b_i)$},
where $s_i=s_i(v_i)$ and $b_i=b_i(v_i)$ denote probability densities
of topological variables $v_i$ for the signal and background,
respectively. The variables we use are the opening angle between the
two leptons in the transverse plane $\Delta\varphi_{\mu\mu}$ ($\mu\mu$
channel), $\met$ ($ee$, $e\mu$ channels), hadronic
missing transverse energy ($\met$ not corrected for lepton momenta)
$\xmet$ (all channels), and the minimum angle between a lepton and
$\xmet$\ direction $\Delta\varphi_{l\not E'_T}^{min}$ (all channels).

We consider four Higgs mass points: 115, 135, 155, and 175~GeV. For
each mass, we construct an individual TLD based on variable
distributions for the mass point. We optimize the TLD cut with respect
to the lowest \mbox{$WH\rightarrow WWW^*$}\ production cross section
limit calculated from the expected number of events given
background-only hypothesis. The contributions from SM background
sources ($WZ$ and $ZZ$) are computed based on their theoretical
production cross sections. The shapes of the variable distributions
for instrumental backgrounds are determined from data. For charge
flips, we use events that pass the same selection as the signal
sample, except that leptons are now required to be of unlike
sign. These events are weighted according to the charge flip
probability as a function of the lepton $p_T$. In addition, for muons
the charge mismeasurement implies that the measured $p_T$ is not
related to the original muon $p_T$, so the resulting distributions are
convoluted with the simulated $p_T$ distribution of mismeasured
muons. For QCD, we reverse the likelihood cut for electrons and
isolation cuts for muons.

\begin{figure}
\begin{center}
\includegraphics[width=\linewidth]{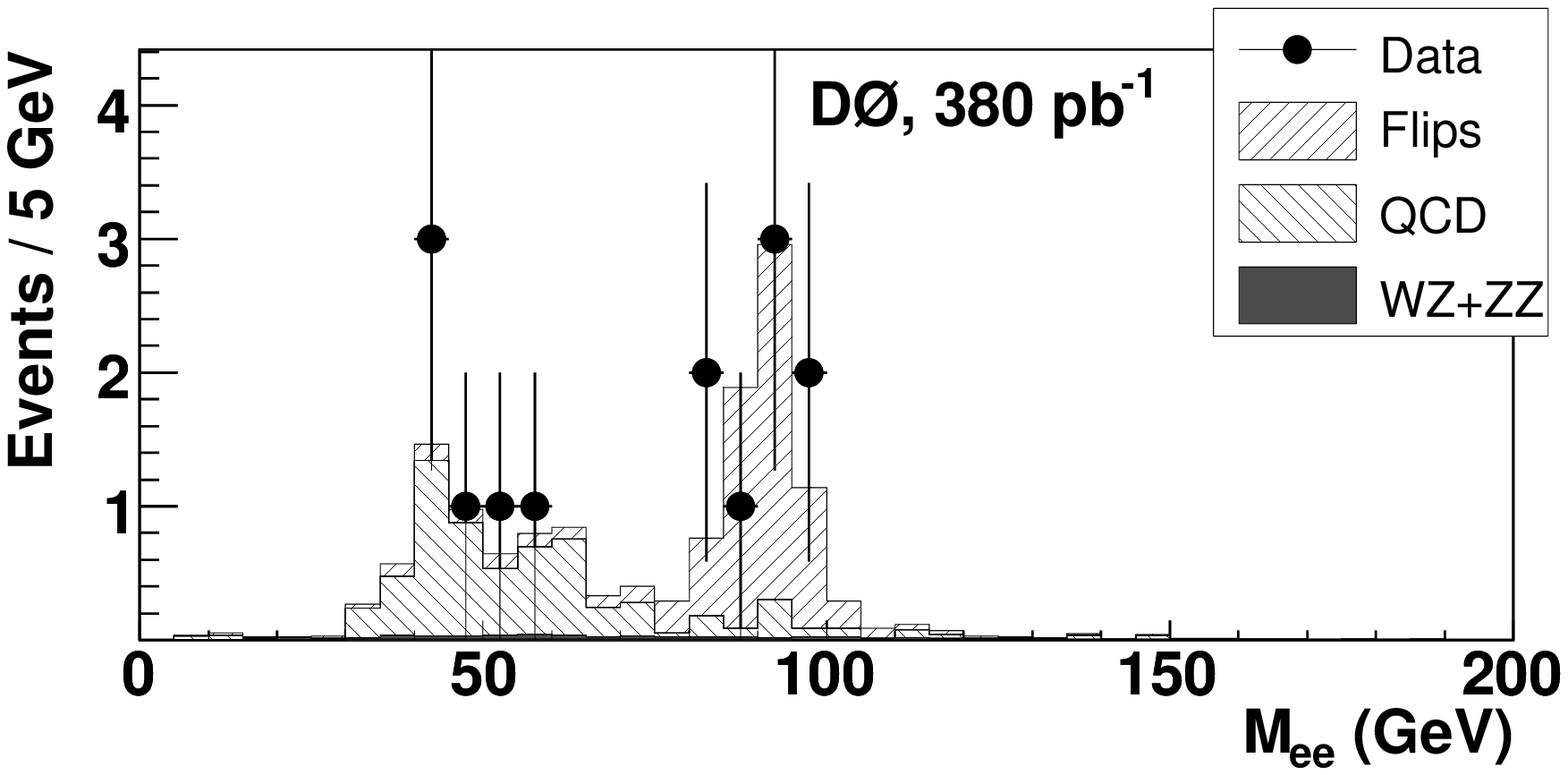}\\
\includegraphics[width=\linewidth]{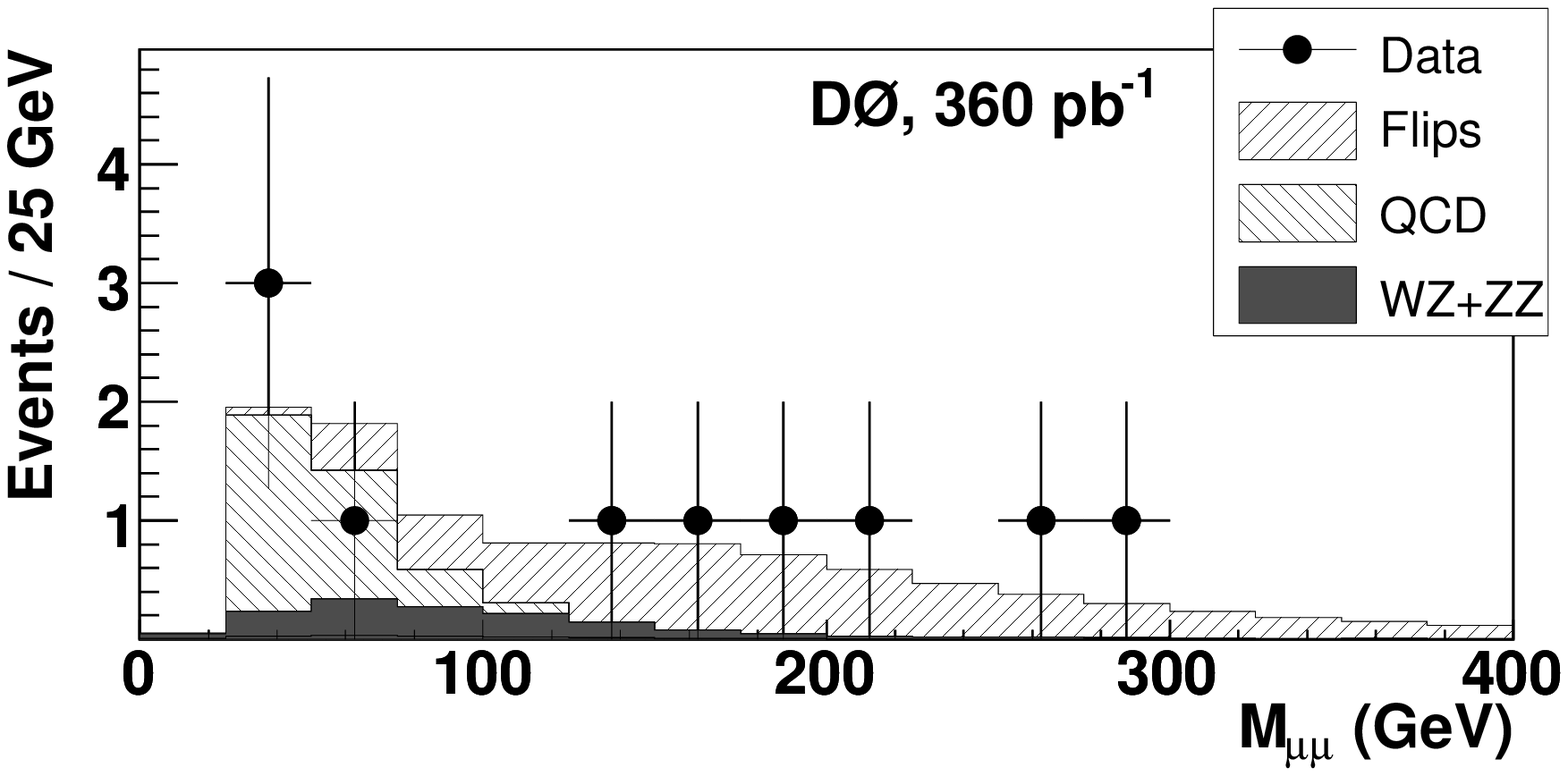}
\end{center}
\caption{\label{fig:mll}
The dilepton invariant mass distribution in the $ee$ (top) and
$\mu\mu$ (bottom) channels fitted to a weighted sum of the
distributions for all backgrounds (for $M_H$=155~GeV).}
\end{figure}

The level of charge flips and QCD background contributions is determined
from the fit of the dilepton invariant mass distribution to a weighted
sum of the distributions for all backgrounds. To avoid potential bias
from the signal, the fit is performed on a sample of events that fail
the TLD cut (``complementary sample''). This procedure is performed
for the $ee$ and $\mu\mu$ channels, as illustrated in
Fig.~\ref{fig:mll}. We verified that the background composition is not
sensitive to the actual value of the TLD cut.
For the $e\mu$ channel, the background due to
charge flips is {\em a priori} small, because $Z/\gamma^*\rightarrow
ll$ production does not contribute to $e\mu$ except via
$Z/\gamma^*\rightarrow\tau\tau\rightarrow e\mu+$neutrinos. The
fraction of charge flips in the $e\mu$ channel is determined from the
charge flip probabilities measured in the $ee$ and $\mu\mu$
channels. The number of background events determined on the
complementary sample is converted to the background expectation in
the signal sample using calculated TLD cut efficiency.

\begin{figure}
\begin{center}
\includegraphics[width=\linewidth]{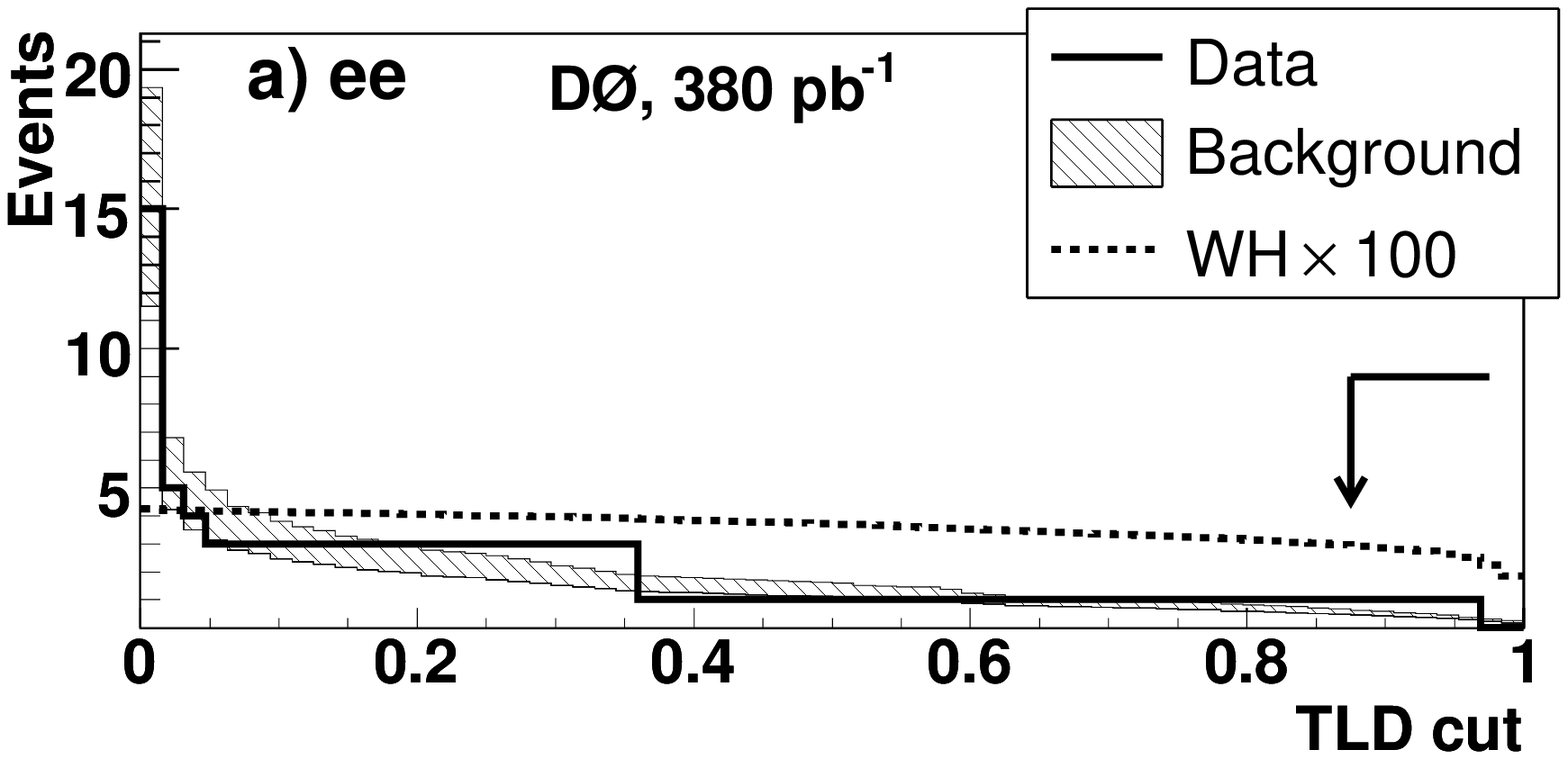}\\
\includegraphics[width=\linewidth]{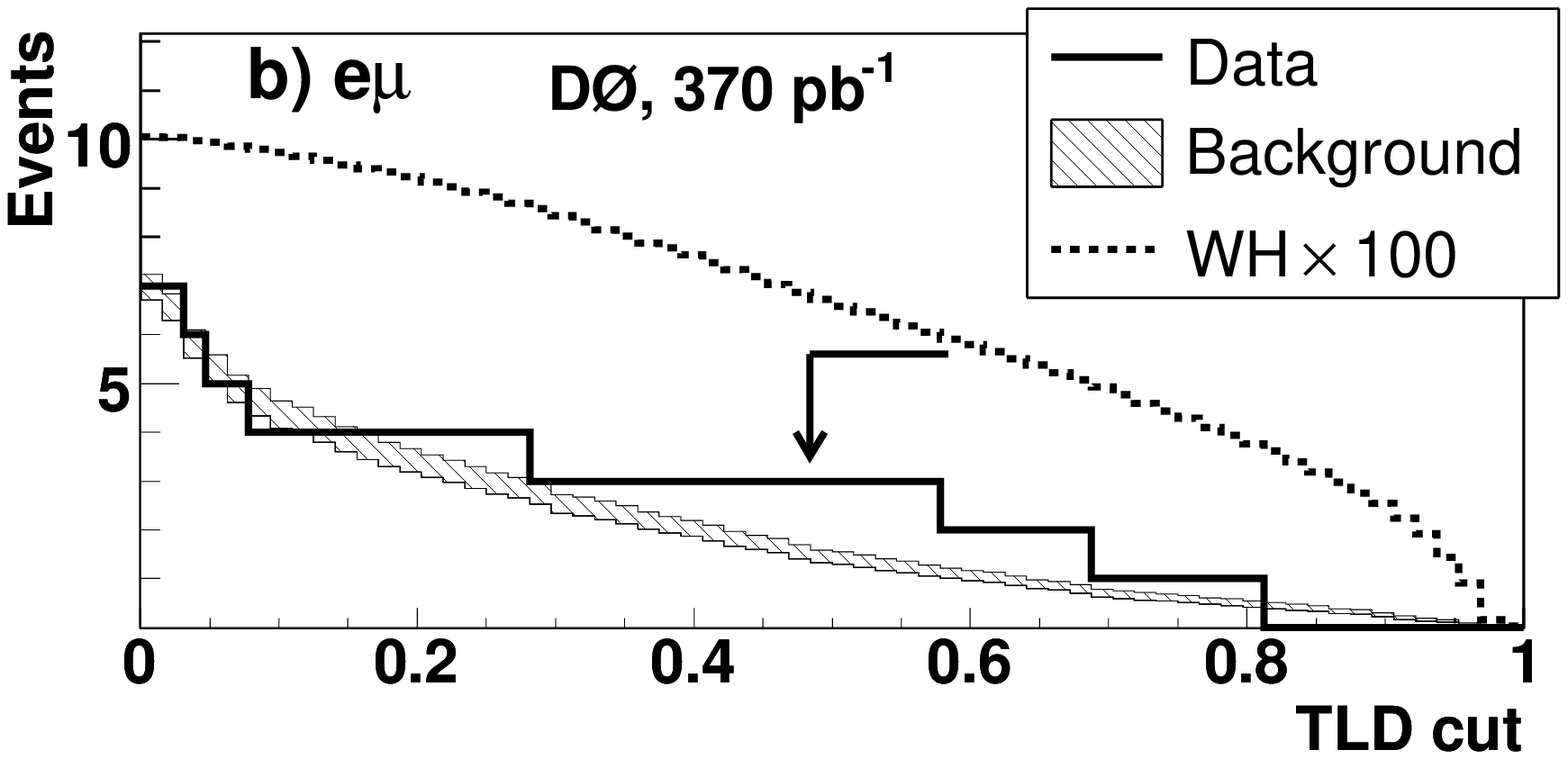}\\
\includegraphics[width=\linewidth]{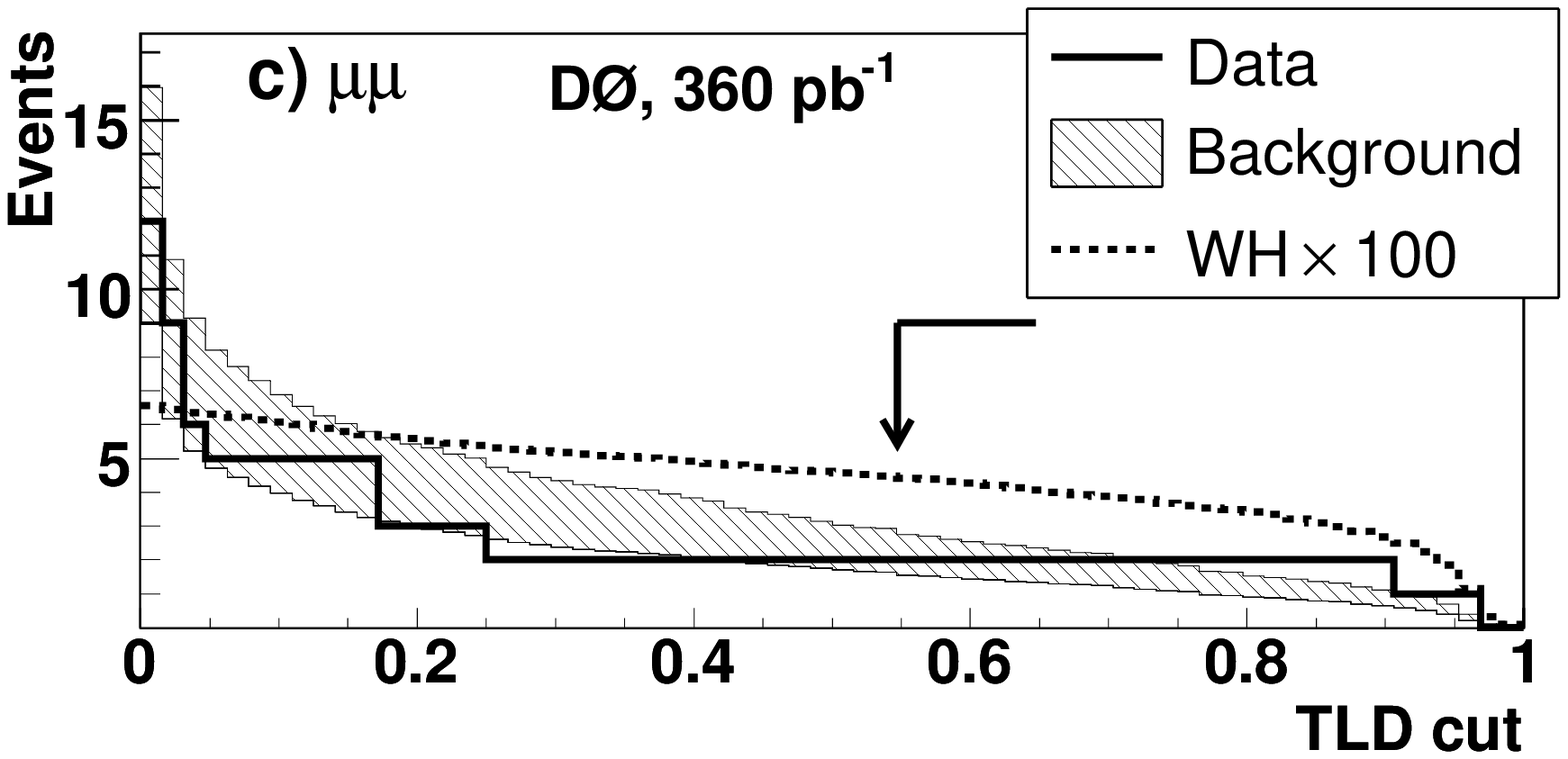}
\end{center}
\caption{\label{fig:lhs}
The observed number of events (solid lines), the predicted background
(shaded bands), and the expected number of signal events times 100
(dashed lines) for $M_H$=155~GeV above the TLD cut in the (a) $ee$,
(b) $e\mu$, and (c) $\mu\mu$ channels. The width of the shaded bands
corresponds to $\pm 1\sigma$ uncertainty on the background
predictions. The vertical arrows indicate the optimal cut values.}
\end{figure}

For all considered Higgs mass points, the number of events remaining
after the TLD cut as a function of the cut value is consistent with
expectations from the SM background, as illustrated in
Fig.~\ref{fig:lhs} for $M_H$=155~GeV. For the optimal TLD cut values,
1 event in the $ee$ channel, 3 events in the $e\mu$ channel, and 2
events in the $\mu\mu$ channel have been observed for each Higgs mass
point of 135, 155, and 175~GeV. For a Higgs mass of 115~GeV, the
observed numbers of events are 1, 4, and 4, respectively, for the
three channels. The number of events after the TLD cut together with
the prediction for the SM background is shown in
Table~\ref{tab:events}.

\begin{table*}
\caption{\label{tab:events}
Number of expected and observed events for a combination of all three
channels after all selections are applied. The errors include both
statistical and systematic uncertainties.}
\begin{ruledtabular}
\begin{tabular}{lcccc}
$M_H$~(GeV) & 115 & 135 & 155 & 175 \\
\hline
Charge flips&  2.35$\pm$0.90 &  1.40$\pm$0.53 &  1.12$\pm$0.43 &  0.89$\pm$0.31 \\
QCD         &  2.35$\pm$1.04 &  2.04$\pm$0.83 &  1.64$\pm$0.69 &  1.16$\pm$0.46 \\
$WZ$        &  3.40$\pm$0.28 &  1.87$\pm$0.15 &  1.51$\pm$0.12 &  1.26$\pm$0.10 \\
$ZZ$        &  0.34$\pm$0.03 &  0.21$\pm$0.02 &  0.17$\pm$0.01 &  0.15$\pm$0.01 \\
\hline
Total       &  8.44$\pm$1.37 &  5.52$\pm$0.99 &  4.45$\pm$0.82 &  3.46$\pm$0.57 \\
\hline
Signal       &  0.037$\pm$0.004 & 0.100$\pm$0.010 & 0.143$\pm$0.015 & 0.110$\pm$0.011 \\
\hline
Data         & 9 & 6 & 6 & 6 \\
\end{tabular}
\end{ruledtabular}
\end{table*}

Limits on the \mbox{$WH\rightarrow WWW^*$}\ production cross section
are calculated using a ``modified frequentist'' approach described in
Ref.~\cite{cls}. The 95\% C.L. limit is defined as the cross section
at which the ratio of the confidence level for the sum of signal and
background hypothesis, $CL_{S+B}$, to the confidence level for the
background to represent the data, $CL_B$, reaches 0.05. The numbers
of observed and expected events in the three channels are input
separately to improve the sensitivity. The uncertainties on the
expected numbers of signal and background events are determined from
the statistical and systematic uncertainties and the luminosity
uncertainty of 6.5\%~\cite{d0lumi}. The signal uncertainty is
10--11\%, depending on the Higgs mass point. The main sources of the
signal uncertainty are the lepton identification, 8\%, and the trigger
efficiency, (4--5)\%. The background uncertainty is 16--18\%,
dominated by the uncertainty on the composition of the instrumental
background, which in turn is mostly due to the limited statistics
of the complementary sample.

\begin{figure}
\includegraphics[width=\linewidth]{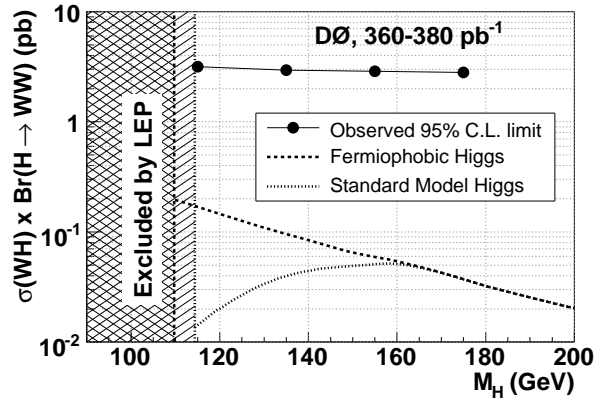}
\caption{\label{fig:limits}
The observed upper limits for the four mass points along with the
theoretical predictions for the SM and fermiophobic Higgs boson
production. Shaded areas correspond to the LEP limits for the SM
(114.4~GeV)~\cite{lepsm} and fermiophobic (109.7~GeV)~\cite{lepff}
Higgs boson.}
\end{figure}

The expected and observed upper limits for the combination of all three
channels are given in Table~\ref{tab:limits}. Figure~\ref{fig:limits}
shows the observed upper limits together with theoretical predictions
for a SM and a fermiophobic Higgs boson. No region can be excluded with the
present data set.

\begin{table}
\caption{\label{tab:limits}Expected and observed upper limits
at the 95\% C.L. for the associated Higgs boson production cross
section times branching fraction
\mbox{$\sigma(WH)\times Br(H\rightarrow WW^*)$}
for various values of $M_H$.}
\begin{ruledtabular}
\begin{tabular}{lcccc}
$M_H$~(GeV) & 115 & 135 & 155 & 175 \\
\hline
Expected limits (pb) & 3.3 & 2.8 & 2.3 & 2.0 \\
Observed limits (pb) & 3.2 & 2.9 & 2.9 & 2.8 \\
\end{tabular}
\end{ruledtabular}
\end{table}

In conclusion, a search has been performed for the process\whprocess\
in the $ee$, $e\mu$, and $\mu\mu$ channels. In all cases the number of observed events is in
agreement with the predicted SM background. The upper limits set on
\mbox{$\sigma(WH)\times Br(H\rightarrow WW^*)$}
for the combination of all three channels vary from 3.2 to 2.8~pb as
the Higgs mass varies from 115 to 175~GeV. In the case of the
fermiophobic Higgs boson with a mass of 115~GeV, this represents a factor
2.4 improvement with respect to the previous D0 result obtained in
Run~I in the $H\rightarrow\gamma\gamma$ decay mode~\cite{d0hgg}. That
becomes a factor 22 improvement for a Higgs mass of 155~GeV.

%
We thank the staffs at Fermilab and collaborating institutions, 
and acknowledge support from the 
DOE and NSF (USA);
CEA and CNRS/IN2P3 (France);
FASI, Rosatom and RFBR (Russia);
CAPES, CNPq, FAPERJ, FAPESP and FUNDUNESP (Brazil);
DAE and DST (India);
Colciencias (Colombia);
CONACyT (Mexico);
KRF and KOSEF (Korea);
CONICET and UBACyT (Argentina);
FOM (The Netherlands);
PPARC (United Kingdom);
MSMT (Czech Republic);
CRC Program, CFI, NSERC and WestGrid Project (Canada);
BMBF and DFG (Germany);
SFI (Ireland);
The Swedish Research Council (Sweden);
Research Corporation;
Alexander von Humboldt Foundation;
and the Marie Curie Program.
%


\begin{thebibliography}{99}
\bibitem{etadef}
The pseudorapidity $\eta=-\ln[\tan(\vartheta/2)]$, where
$\vartheta$ is the polar angle.
The D0 coordinate system has the $z$~axis along the Tevatron proton beam
direction.
\bibitem{drdef}
The distance $\Delta R$ is defined as
$\Delta R=\sqrt{(\Delta\varphi)^2+(\Delta\eta)^2}$, where $\varphi$ is
the azimuthal angle.
\end{thebibliography}
\end{document}